\documentclass[11pt]{article}
\usepackage{amsmath}
\usepackage{graphics}

\def\be{\begin{equation}}
\def\ee{\end{equation}}
\def\ed{\end{document}}
\def\vs1{{\vskip .3cm}}
\input{tcilatex}

\begin{document}

\title{Relativistic wave equations of $n$-body systems of fermions and antifermions
of various masses in quantum electrodynamics}
\author{Mohsen Emami-Razavi$^{1}$, Nantel Bergeron$^{2}$ and Jurij W. Darewych$^{3}$ 
\\
$^{1}$ Centre for Research in Earth and Space Science, \\
$^{2}$ Department of Mathematics and Statistics, \\
$^{3}$ Department of Physics and Astronomy, \\
York University, Toronto, Ontario, M3J~1P3, Canada}
\date{17 Sept 2012}
\maketitle

\begin{abstract}
The variational method in a reformulated Hamiltonian formalism of Quantum
Electrodynamics is used to derive relativistic wave equations for systems
consisting of $n$ fermions and antifermions of various masses. The derived
interaction kernels of these equations include one-photon exchange
interactions. The equations have the expected Schr\"odinger non-relativistic
limit. Application to some exotic few lepton systems is discussed briefly.
\end{abstract}




\section{Introduction:}

It is straightforward to write down the non-relativistic (Schr\"{o}dinger)
equation for a system of $n$ particles (fermions or bosons), particularly
since the interparticle interactions can be adequately described by
potentials. Such is not the case for relativistic systems, and a many-body
quantized field theory must be used in which the quanta of the interactions
are treated on par with the particles. In a recent paper, M. Emami-Razavi
derived relativistic $n$-body wave equations of combined fermions and
antifermions with equal masses, including the interactions, starting from
the underlying Quantum Electrodynamics (QED) Lagrangian. \cite{MohsenPRA}. A
generalization to systems of $n$ fermions and antifermions of different
masses is presented in this paper.

Many body fermion systems with electromagnetic interactions are the
substance of atomic physics. Fundamental fermion and antifermion systems
with electromagnetic interactions are of particular interest because they
are ``pure'' QED systems, with point-like constituents and no nuclear force
or size effects. Examples of such systems include positronium (Ps: $%
e^{+}e^{-}$) and muonium (Mu: $\mu ^{+}e^{-}$) and their ions (Ps$^{-}$: $%
e^{+}e^{-}e^{-}$, Mu$^{-}$: $\mu ^{+}e^{-}e^{-}$), four body systems such as
Ps$_{2}$ ($e^{+}e^{-}e^{+}e^{-}$), Mu$_{2}$ ($\mu ^{+}e^{-}\mu ^{+}e^{-}$).
More generally, for ``exotic atoms'' such as $e^{+}$PsH, and Ps$_{2}$O, Li$%
^{+}$Ps$_{2}$ or Na$^{+}$Ps$_{2}$, etc., nuclear size effects are not
negligible. These have received attention in the past (\textit{e.g.} ref. 
\cite{Mitroy2002}).

The problem of describing relativistic bound states in quantum field theory
(QFT) was solved many years ago by Bethe and Salpeter (BS) \cite{BS}-\cite
{BS1}, at least in principle. However, the BS method is not free of
complications, such as the existence of relative-time coordinates,
difficulty of implementation for systems of more than two bodies, and in
practice, the perturbative treatment of interactions. There are many papers
available in the literature that use the BS method\textbf{, }at least for
two and three body systems. For example, Adkins and co-workers \cite{Adkins1}%
-\cite{AD1} have used Bethe-Salpeter formalism for the calculation of recoil
corrections to the energy levels of hydrogenic ions, and a discussion of
issues that will have to be treated for the many-electron case, where highly
accurate experiments have been carried out, is given. Using a different
approach than the BS formalism, Barut \cite{Barut} summarized his previous
work (including work with his\textbf{\ }co-authors) and generalized his
two-body QED equation to many-body particles interacting via the exchange of
massless vector bosons. In his formulation \cite{Barut}, the relativistic
many-body problem has a structure that is similar to that of the
Schr\"{o}dinger many-body problem.

An alternative to the BS and other approaches is the variational method
within the reformulated Hamiltonian formalism of QFT, introduced by Darewych 
\cite{DarLdeB}-\cite{Dar1}. Among the appealing features of this approach is
that it is straightforwardly generalizable to systems of more than two
particles, and it can be cast in the form of a relativistic generalization
of the Schr\"{o}dinger description of $n$-body systems. As a variational
method it is applicable, at least in principle, to strongly coupled systems
for which perturbation theory may be unreliable. It has disadvantages as
well, particularly in that it may not be manifestly covariant, and like all
variational methods, the construction of realistic yet tractable trial
states may be a difficult task. A variational approximation is, of course,
only as good as the trial states that are being employed. In this paper we
use the formalism of \cite{DarLdeB}-\cite{Dar1} to derive relativistic wave
equations for systems of $n$ fermions and antifermions of different mass
(where $n$ can be any integer number). To our knowledge, there are no papers
in the literature that present relativistic wave equations for a system of $%
n $ fermion with different masses in QED.

Regarding relativistic or QED corrections \textbf{t}o the non-relativistic
eigenenergies of few body exotic systems we can mention the following.
Accurate calculations of the positronium hyperfine structure, i.e. the $%
O(\alpha ^{6})$ contributions to ground-state hyperfine splitting in
positronium\textbf{,} have been studied by Adkins et al. \cite{Adkins}-\cite
{A4}. Theoretical studies of the Ps$^{-}$ ($e^{-}e^{+}e^{-}$) are now well
advanced, including perturbative determinations of relativistic and QED
corrections (cf. Drake and Grigorescu \cite{drake3} and references
therein.); the leading relativistic and QED corrections to the
nonrelativistic ground state energy of the three-body system Ps$^{-}$ have
been calculated numerically using a Hylleraas correlated basis set. The
corrections to this energy in ref. \cite{drake3} include the lowest order
Breit interaction, the vacuum polarization potential, one and two photon
exchange contributions, the annihilation interaction, and spin-spin contact
terms. In a recent work, Bubin et al. \cite{Pos10} reported that they have
obtained a very accurate variational wave function for non-relativistic
binding energy of the positronium molecule (Ps$_{2}$, $e^{+}e^{-}e^{+}e^{-}$%
), which they used to calculate the relativistic corrections.

Experiments on such exotic atoms, though difficult, have been and are being
undertaken not only for positronium, muonium and the three-body Ps$^{-}$
system, but also for the four-body ``positronium molecule'' (Ps$_{2}$: $%
e^{+}e^{-}e^{+}e^{-}$). In 2007, the positronium molecule was observed by
Cassidy and Mills \cite{Cassidy}. Ps$_{2}$ creation \cite{Cassidy} heralds a
new chapter in the study of matter and antimatter and the binding of
positrons to ordinary atoms \cite{Mitroy2002}. 
Another example is the system consisting of three distinct fermions\textbf{\
($e^{-},\tau ^{-},\mu ^{+}$) }\cite{Mu08}. To our knowledge this system has
not been observed yet. Other systems of interest are four or more fermions
interacting electromagnetically, for example, the system ($m^{Z^{+}},$Ps$%
^{-} $), where $m$ is the mass of a specified particle \cite{Mitroy2004}. 
The properties of some exotic five-particle systems have been studied in 
\cite{Mitroy2001}. %
%

In general, the approach to the calculation of the properties of systems
like those cited in the preceding paragraphs has been to calculate accurate
solutions of the non-relativistic (Schr\"{o}dinger) equation then to use
these to calculate relativistic and QED corrections to the non-relativistic
eigenenergies by means of perturbation theory. We present a method in which
the starting point is from relativistic equations that include all
``tree-level'' interactions. As will be explained below, this approach
automatically includes all effects to $O(\alpha ^{4})$, much like the Dirac
equation for the one-electron system in a Coulomb potential, which yields
eigenenergies that are correct to $O(\alpha ^{4})$.

The presentation of this paper is as following. The relativistic $n$-body
QED wave equations for different ``flavors'' are presented in section II.
Two, three, and four-body examples are given in section III. Concluding
remarks are presented in section IV.

%
%


\section{QED $n$-body wave equations:}

As in previous work \cite{MohsenPRA}, we use the variational method in the
reformulated version of QED (Darewych \cite{DarLdeB}, Terekidi and Darewych 
\cite{TD1}), in which Green's functions are used to express the mediating
field in terms of the particle fields. %

We start with the usual Lagrangian (density) for $n$ fermions and
antifermions, corresponding to fields $\psi _{j}(x)$ and masses $m_{j}$ ($%
j=1,$ $2,$ $3,...$), interacting electromagnetically ($\hbar =c=1)$: 
\begin{equation}
\mathcal{L}=\underset{j=1...n}{\sum }\overline{\psi }_{j}(x)\left( i\gamma
^{\mu }\partial _{\mu }-m_{j}-q_{j}\gamma _{\mu }A^{\mu }(x)\right) \psi
_{j}(x)-\frac{1}{4}F^{\mu \nu }(x)F_{\mu \nu }(x),  \label{1.1}
\end{equation}
%
where $x=(t,\mathbf{r})$, $A^{\mu }(x)$ is the photon field, $F_{\mu \nu
}(x)=\partial _{\mu }A_{\nu }(x)-\partial _{\nu }A_{\mu }(x)$, and the
coupling constants $q_{j}$ can have arbitrary values. 
The equations of motion that follow from (\ref{1.1}) are 
\begin{equation}
\left( i\gamma ^{\mu }\partial _{\mu }-m_{j}\right) \psi _{j}(x)=q_{j}\gamma
^{\mu }A_{\mu }(x)\psi _{j}(x),  \label{1.2}
\end{equation}
and 
\begin{equation}
\partial _{\mu }F^{\mu \nu }(x)=j^{\nu }(x),  \label{1.3}
\end{equation}
where 
\begin{equation}
j^{\nu }(x)=\underset{j=1...n}{\sum }q_{j}\,{\overline{\psi }}_{j}(x)\gamma
^{\nu }\psi _{j}(x).  \label{2}
\end{equation}
As is well known, the Maxwell equation (\ref{1.3}) has the formal solution 
\begin{equation}
A_{\mu }(x)=A_{\mu }^{0}(x)+\int d^{4}x^{\prime }D_{\mu \nu }(x-x^{\prime
})j^{\nu }(x^{\prime }),  \label{1.4}
\end{equation}
where $A_{\mu }^{0}(x)$ is a solution of the homogeneous (``free field'')
equation (\ref{1.3}) (\textit{i.e.\ } with $j^{\nu }(x)=0$) and $D_{\mu \nu
}(x-x^{\prime })$ is the a Green function (or ``photon propagator''), such
that 
\begin{equation}
\partial _{\alpha }\partial ^{\alpha }D_{\mu \nu }(x-x^{\prime })-\partial
_{\mu }\partial ^{\alpha }D_{\alpha \nu }(x-x^{\prime })=g_{\mu \nu }\delta
^{4}(x-x^{\prime }).  \label{1.5}
\end{equation}
Substitution of the formal solution (\ref{1.4}) into (\ref{1.2}) yields the
result 
\begin{equation}
\left( i\gamma ^{\mu }\partial _{\mu }-m_{j}\right) \psi _{j}(x)=q_{j}\gamma
^{\mu }\left( A_{\mu }^{0}(x)+\int d^{4}x^{\prime }D_{\mu \nu }(x-x^{\prime
})j^{\nu }(x^{\prime })\right) \psi _{j}(x).  \label{1.6}
\end{equation}
Equations (\ref{1.6}) are derivable from the stationary action principle $%
\displaystyle\delta \int d^{4}x\,\mathcal{L}_{R}(x)=0$, where 
\begin{equation}
\mathcal{L}_{R}=\underset{j=1...n}{\sum }\overline{\psi }_{j}(x)\left(
i\gamma ^{\mu }\partial _{\mu }-m_{j}-q_{j}\gamma ^{\mu }A_{\mu
}^{0}(x)\right) \psi _{j}(x)-\frac{1}{2}\int d^{4}x^{\prime }j^{\mu
}(x^{\prime })D_{\mu \nu }(x-x^{\prime })j^{\nu }(x),  \label{1}
\end{equation}
provided that the Green function $D_{\mu \nu }(x-x^{\prime })$ is symmetric.

The Hamiltonian density corresponding to the Lagrangian (\ref{1}) is
obtained using the usual canonical prescription 
\begin{equation}
{\mathcal{H}}=\sum_{j}\pi _{\psi _{j}}\,{\dot{\psi}_{j}}-{\mathcal{L}}%
_{R},~~~~\mathrm{where}~~~\pi _{\psi _{j}}=\frac{\partial \mathcal{L}_{R}}{%
\partial {\dot{\psi}_{j}}}=i\psi _{j}^{\dagger }.  \label{411}
\end{equation}
This yields the result 
\begin{equation}
\mathcal{H}(x)=\mathcal{H}_{0}(x)+\mathcal{H}_{I_{1}}(x)+\mathcal{H}%
_{I_{2}}(x),  \label{5}
\end{equation}
where 
\begin{equation}
\mathcal{H}_{0}(x)=\underset{j=1...n}{\sum }\psi _{j}^{\dagger }(x)~(-i%
\overset{\rightarrow }{\alpha }\cdot \nabla +m_{j}\beta )~\psi _{j}(x),
\label{6}
\end{equation}
\begin{equation}
\mathcal{H}_{I_{1}}(x)=\underset{j=1...n}{\sum }q_{j}\,{\overline{\psi }}%
_{j}(x)\,\gamma ^{\mu }\,A_{\mu }^{0}(x)\psi _{j}(x),  \label{7}
\end{equation}
\begin{equation}
\mathcal{H}_{I_{2}}(x)=\frac{1}{2}\int d^{4}x^{\prime }j^{\mu }(x^{\prime
})D_{\mu \nu }(x-x^{\prime })j^{\nu }(x),  \label{8}
\end{equation}
with $j^{\nu }(x)=\underset{j=1...n}{\sum }q_{j}{\overline{\psi }}%
_{j}(x)\gamma ^{\nu }\psi _{j}(x)$ and $\displaystyle D_{\mu \nu
}(x-x^{\prime })=\int \frac{1}{(2\pi )^{4}}d^{4}k\,D_{\mu \nu }(k)\,\exp
(-ik(x-x^{\prime }))$. %

\vskip.2cm We shall consider only stationary bound-states of $n$-fermion
systems (or their scattering) in this work. Thus we shall not consider
processes that involve the emission or absorption of free (physical) photons
in the present work. Moreover, the variational trial states that we use (see
eqs. (\ref{17}) and (\ref{18})) do not sample the term corresponding to eq. (%
\ref{7}), \textit{i.e} the matrix elements of $H_{I_{1}}$ with respect to
the trial states (\ref{17}) and (\ref{18})) vanish. Consequently we shall
leave out the free photon field $A_{\mu }^{0}(x)$ in what follows. Thus, the
Hamiltonian density that we shall work with, namely $\mathcal{H}(x)=\mathcal{%
H}_{0}(x)+\mathcal{H}_{I_{2}}(x)$, contains the Fermion fields and the
photon propagators $D_{\mu \nu }$ only. Nevertheless, one should note that
virtual photons are included in our work, by virtue of the photon propagator
in the interaction term, eq. (\ref{8}).

In practice, a choice of gauge is needed to specify the Green function. The
covariant Lorentz gauge $\partial _{\mu }A^{\mu }=0$ is a frequent choice.
For bound-state problems the Coulomb gauge is a convenient choice. In
momentum representation it is 
\begin{equation}
D_{00}(k)=\frac{1}{|\mathbf{k|}^{2}},~~~D_{0j}(k)=0,~~~D_{ij}(k)=\frac{1}{%
k^{\mu }k_{\mu }}\left( \delta _{ij}-\frac{k_{i}\,k_{j}}{\mathbf{k}^{2}}%
\right) .  \label{9}
\end{equation}

We construct a quantum field theory, based on the reformulated Hamiltonian,
by promoting the field variables to operators which satisfy the usual
(anti)commutation rules. 
Furthermore, we use the usual Fourier decomposition 
\begin{equation}
\psi _{j}(x)=\sum_{s}\int \frac{d^{3}p}{\left( 2\pi \right) ^{3/2}}\left( 
\frac{m_{j}}{\omega _{jp}}\right) ^{1/2}\left[ b_{_{j}}\left( \mathbf{p}%
,s\right) u_{j}\left( \mathbf{p},s\right) e^{-ip\cdot x}+d_{j}^{\dagger
}\left( \mathbf{p},s\right) v_{j}\left( \mathbf{p},s\right) e^{ip\cdot x}%
\right] ,  \label{11}
\end{equation}
where $p=p^{\mu }=\left( \omega _{jp},\mathbf{p}\right) $, and $\omega _{jp}=%
\sqrt{m_{j}^{2}+\mathbf{p}^{2}}$; that is, the field operators $\psi_j$ (and
their conjugates $\psi^{\dagger}_j)$ are replaced by linear superpositions
of ``ladder" operators $b_j, d^{\dagger}_j$ (and their conjugates $%
b^{\dagger}_j, d_j)$. The mass-$m_{j}$ free-particle Dirac spinors $u_{j}$\
and $v_{j}$,\ where $\left( \gamma ^{\mu }{p}_{\mu }-m_{j}\right)
u_{j}\left( \mathbf{p},s\right) =0$,\ $\left( \gamma ^{\mu }{p}_{\mu
}+m_{j}\right) v_{j}\left( \mathbf{p},s\right) =0$, are normalized such that 
\begin{equation}
u_{j}^{\dagger }\left( \mathbf{p},s\right) u_{j}\left( \mathbf{p},\sigma
\right) =v_{j}^{\dagger }\left( \mathbf{p},s\right) v_{j}\left( \mathbf{p}%
,\sigma \right) =\frac{\omega _{jp}}{m_{j}}\delta _{s\sigma }, ~~\mathrm{with%
} ~~ u_{j}^{\dagger }\left( \mathbf{p},s\right) v_{j}\left( \mathbf{p}%
,\sigma \right) =v_{j}^{\dagger }\left( \mathbf{p},s\right) u_{j}\left( 
\mathbf{p},\sigma \right) =0.  \label{13}
\end{equation}
The creation and annihilation operators $b^{\dagger }$, $b$ of the (free)
fermions of mass $m_{j}$, and $d^{\dagger }$, $d$ for the corresponding
antiparticles, satisfy the usual anticommutation rules. The non-vanishing
ones are 
\begin{equation}
\left\{ b_{_{j}}\left( \mathbf{p},s\right) ,b_{j}^{\dagger }\left( \mathbf{q}%
,\sigma \right) \right\} =\left\{ d_{j}\left( \mathbf{p},s\right)
,d_{j}^{\dagger }\left( \mathbf{q},\sigma \right) \right\} =\delta _{s\sigma
}\delta ^{3}\left( \mathbf{p}-\mathbf{q}\right) .  \label{14}
\end{equation}
The vacuum state $|0\rangle $ is defined by $b_{_{j}}\left( \mathbf{p}%
,s\right) |0\rangle =d_{j}(\mathbf{p,}s)|0\rangle =0$. Note that this is a
``filled negative-energy sea" vacuum, as is clear from the fact that the
eigenvalues of ${\hat H}_0$ are positive and there are no ``negative-energy"
solutions.

With the replacement (\ref{11}), and its conjugate for $\psi _{j}^{\dagger }$%
, the Hamiltonian operator, $\displaystyle{\hat{H}}={\int d^{3}x\;{\hat{%
\mathcal{H}}}(x)}$ is expressed in terms of the creation and annihilation
operators $b^{\dagger },d^{\dagger }$, $b,d$ in the usual way. Note that we
normal-order the entire Hamiltonian (thereby denoting it :${\hat{H}}$:),
since this circumvents the need for mass renormalization which would
otherwise arise. Not that there is a difficulty with handling mass
renormalization in the present formalism, as shown in various earlier papers
(see, for example, \cite{TD1}). It is simply that we are not interested in
mass renormalization here, since it has no effect on bound state energies.
Furthermore, the approximate trial states which we use in this work, are
incapable of sampling loop effects. Thus, the normal ordering of the entire
Hamiltonian does not sweep under the carpet loop renormalization effects,
since none arise at the present level of approximation.

%
The Hamiltonian operator, specifically its interaction part ${\hat{H}}%
_{I_{2}}$, is time dependent (the free-field part, ${\hat{H}}_{0}$, is
time-independent). For the description of stationary states it is convenient
to work with a time-independent Hamiltonian. This can be achieved by a
unitary transformation of states and operators by means of the unitary
operator $\displaystyle e^{iH_{0}t}$. The result is the same as setting $t=0$
in all matrix elements below, which we henceforth do. We might mention that
if we do not set $t=0$, we would have extra factors of the form the of $e^{i%
\mathbf{(}\omega _{jp_{j}^{\prime }}+\omega _{kp_{k}^{\prime }}-\omega
_{jp_{j}}-\omega _{kp_{k}})t}$ in the matrix elements of $H_{I_{2}}$; but,
as already stated, this ``phase factor'' time dependence can be eliminated
(rotated away to unity) by a unitary transformation or, equivalently,
setting $t=0$.

%

In the Hamiltonian formalism of QFT we seek solutions of the eigenvalue
equation. 
\begin{equation}
{\hat{P}}^{\beta }|\Psi {\rangle }=Q^{\beta }|\Psi {\rangle },  \label{15}
\end{equation}
where ${\hat{P}}^{\beta }=({\hat{H}},{\hat{\mathbf{P}}})$ is the
energy-momentum operator of the QFT, and $Q^{\beta }=(E,\mathbf{Q})$ is the
energy-momentum eigenvalue. The case $\mathbf{Q}=0$ defines the rest-frame
of the system. The $\beta =0$ component of (\ref{15}) is generally not
solvable, hence approximation methods, such as the variational method, must
be used. The latter amounts to finding approximate solutions by using the
variational principle 
\begin{equation}
\delta \langle \Psi_{tr} |:{\hat{H}}-E:|\Psi_{tr} \rangle _{t=0}=0,
\label{16}
\end{equation}
where $|\Psi_{tr} \rangle $ is a suitably chosen trial state.

%
%


We consider two explicit types of $n$-fermion systems, one with an equal
number of particles and antiparticles and one with the number of particles
greater by one that the number of antiparticles. For a system of $%
\displaystyle\frac{n}{2}$ particles and $\displaystyle\frac{n}{2}$
antiparticles (here $n$ is even) the simplest trial state is 
\begin{equation}
|\Psi _{n}\rangle =\underset{s_{1}...s_{n}}{\sum }\int
d^{3}p_{1}...d^{3}p_{n}\,F_{s_{1}s_{2}...s_{n}}(\mathbf{p}_{1},...,\mathbf{p}%
_{n})\,b_{1}^{\dagger }(\mathbf{p}_{1},s_{1})d_{2}^{\dagger }(\mathbf{p}%
_{2},s_{2})....b_{n-1}^{\dagger }(\mathbf{p}_{n-1},s_{n-1})d_{n}^{\dagger }(%
\mathbf{p}_{n},s_{n})|0\rangle ,~  \label{17}
\end{equation}
where the functions $F_{s_{1}...s_{n}}(\mathbf{p}_{1},...,\mathbf{p}_{n})$
are well-behaved, adjustable function (normalizable for bound states); $%
s_{j},\,\mathbf{p}_{j}(j=1,..n)$ are the spin indices and momentum
coordinates of the fermions that make up the $n$-body system. Similarly, for
a system of $\displaystyle\frac{n+1}{2}$ particles and $\displaystyle\frac{%
n-1}{2}$ antiparticles (here $n$ is odd) we have 
\begin{equation}
|\Psi _{n}\rangle =\underset{s_{1}...s_{n}}{\sum }\int
d^{3}p_{1}...d^{3}p_{n}\,F_{s_{1}...s_{n}}(\mathbf{p}_{1},...,\mathbf{p}%
_{n})\,b_{1}^{\dagger }(\mathbf{p}_{1},s_{1})d_{2}^{\dagger }(\mathbf{p}%
_{2},s_{2})....d_{n-1}^{\dagger }(\mathbf{p}_{n-1},s_{n-1})b_{n}^{\dagger }(%
\mathbf{p}_{n},s_{n})|0\rangle .~  \label{18}
\end{equation}
The variational coefficient functions $F_{s_{1}...s_{n}}(\mathbf{p}_{1},...,%
\mathbf{p}_{n})$ will be determined in accordance with the variational
principle (\ref{16}). Not all of them will be independent since they must be
chosen to be eigenstates of the total momentum and angular momentum
(magnitude and projection) of the field theory.

The trial states (\ref{17}) and (\ref{18}) 
are variational approximations to the unknown exact eigenstates of the
Hamiltonian. They are constructed as superpositions of the eigenstates of
the free Hamiltonian $H_{0}$, equation (\ref{6}), the so-called
``Fock-states''. This is analogous to expanding an eigenfunction $\phi (%
\mathbf{r})$ of a one-particle system in Schr\"{o}dinger quantum mechanics
in terms of the free-particle states\ $e^{i\mathbf{p\cdot r}}$, i.e.\ $\phi (%
\mathbf{r})$\ $=\int d^{3}p$ $f(\mathbf{p}) \, e^{i\mathbf{p.r}}$ (which
would be the exact eigensolution with an appropriate choice of $f(\mathbf{p}%
) $). Of course, in QFT the state (\ref{17}) or (\ref{18}) cannot be an
exact eigenstate of the full Hamiltonian, no matter what\textbf{\ }choice of%
\textbf{\ }$F_{s_{1}...s_{n}}(\mathbf{p}_{1},...,\mathbf{p}_{n})$\textbf{\ }%
is made. Rather, they are approximate (or ``trial'') states. %
%

To implement the variational principle (\ref{16}) we must calculate the
matrix elements of the Hamiltonian operator using the trial states (\ref{17}%
), (\ref{18})\textbf{.} The matrix element corresponding to the
rest-plus-kinetic energy of such a $n$-fermion system is 
\begin{equation}
\langle \Psi _{n}|:{\hat{H}}_{_{\psi }}-E:|\Psi _{n}\rangle =\underset{%
s_{1}...s_{n}}{\sum }\int d^{3}p_{1}...d^{3}p_{n}\,F_{s_{1}...s_{n}}^{\ast }(%
\mathbf{p}_{1},...,\mathbf{p}_{n})F_{s_{1}...s_{n}}(\mathbf{p}_{1},...,%
\mathbf{p}_{n})\big[\omega _{1p_{1}}+\cdots +\omega _{np_{n}}-E\big].
\label{19}
\end{equation}
The matrix element corresponding to the interactions is a sum of terms
corresponding to attractive one photon exchange (plus repulsive virtual
annihilation interactions for each particle-antiparticle pair, if there were
any) and repulsive one photon exchange between pairs of fermions of the same
sign of charge. Symbolically, 
\begin{equation}
\langle \Psi _{n}|:{\hat{H}}_{I}:|\Psi _{n}\rangle =[\mathcal{M}^{\text{%
\textsl{Attractive}}}]+[\mathcal{M}^{\text{\textsl{Repulsive}}}].  \label{20}
\end{equation}
If $n$ is even (\textit{i.e.} an equal number of particles and
antiparticles), there are $n^{2}/4$ particle-antiparticle combinations and $%
(n^{2}-2n)/4\ $ two-identical-charged-fermion combinations. For example, for 
$n=4,$ we have 4 attractive one-photon exchange terms, 2 repulsive
one-photon exchange terms. If $n$ is odd (\textit{i.e.} one more particle
than antiparticle) there are $(n^{2}-1)/4$ attractive and $(n-1)^{2}/4$
repulsive terms. For example, for $n=5$, we have 6 attractive, 4 repulsive
terms.

For the $n$-body system described by the trial state (\ref{17}) or (\ref{18}%
), the matrix element corresponding to the interactions is $\langle \Psi
_{n}|:{\hat{H}}_{I}:|\Psi _{n}\rangle =\langle \Psi _{n}|:{\hat{H}}%
_{I_{2}}:|\Psi_{n}\rangle$, since $\langle \Psi _{n}|:{\hat{H}}%
_{I_{1}}:|\Psi_{n}\rangle = 0$. That is, as stated previously, the
variational trial states (\ref{17}) or (\ref{18}) do not sample the term $:{%
\hat{H}}_{I_{1}}:$ (cf. eq. (\ref{7})) of the interaction Hamiltonian. To
repeat, this means that with such simple trial states only stationary
(stable bound or scattering) states can be described, but not processes that
involve the emission or absorption of physical photons. Thus, 
\begin{eqnarray}
\langle \Psi _{n}| &:&{\hat{H}}_{I}\text{ \ }:\text{ }|\Psi _{n}\rangle
=\langle \Psi _{n}|:\text{ \ }{\hat{H}}_{I_{2}}\text{ \ }:\text{ }|\Psi
_{n}\rangle  \notag \\
&=&\underset{s_{1}^{\prime }...s_{n}^{\prime }}{\underset{s_{1}...s_{n}}{%
\sum }}\int d^{3}p_{1}...d^{3}p_{n}~d^{3}p_{1}^{\prime
}...d^{3}p_{n}^{\prime }\,F_{s_{1}^{\prime }s_{2}^{\prime }...s_{n}^{\prime
}}^{\ast }(\mathbf{p}_{1}^{\prime },...\mathbf{p}_{n}^{\prime
})\,F_{s_{1}s_{2}...s_{n}}(\mathbf{p}_{1},...,\mathbf{p}_{n})  \notag \\
&&\Bigg\{{\overset{n-1}{\underset{j=1}{\sum }}}~{{\sum_{k=j+1}^{n}}}{%
\overset{\overset{}{\!\!^{^{\prime }}}}{}}~~\left[ \underset{i=1...n}{\prod }%
^{(j,k)}\delta _{s_{i}^{\prime }s_{i}}\right] \left[ \underset{i=1...n}{%
\prod }^{(j,k)}\delta ^{3}(\mathbf{p}_{i}^{\prime }-\mathbf{p}_{i})\right] 
\frac{m_{j}m_{k}q_{j}q_{k}}{2(2\pi )^{3}}\frac{\delta ^{3}(\mathbf{p}%
_{j}^{\prime }+\mathbf{p}_{k}^{\prime }-\mathbf{p}_{j}-\mathbf{p}_{k})}{%
\sqrt{\omega _{jp_{j}^{\prime }}\omega _{kp_{k}^{\prime }}\omega
_{jp_{j}}\omega _{kp_{k}}}}\times  \notag \\
&&\left( -\mathcal{M}_{s_{j}s_{k}s_{j}^{\prime }s_{k}^{\prime }}^{\text{%
\textsl{Attractive} }}(\mathbf{p}_{j},\mathbf{p}_{k},\mathbf{p}_{j}^{\prime
},\mathbf{p}_{k}^{\prime })\right) ~~{+~~\overset{n-2}{\underset{j=1}{\sum }}%
~{\sum_{k=j+2}^{n}}}{\overset{\overset{}{\!\!^{^{\prime }}}}{}}~~\left[ 
\underset{i=1...n}{\prod }^{(j,k)}\delta _{s_{i}^{\prime }s_{i}}\right] %
\left[ \underset{i=1...n}{\prod }^{(j,k)}\delta ^{3}(\mathbf{p}_{i}^{\prime
}-\mathbf{p}_{i})\right] \times  \notag \\
&&\frac{m_{j}m_{k}q_{j}q_{k}}{2(2\pi )^{3}}\frac{\delta ^{3}(\mathbf{p}%
_{j}^{\prime }+\mathbf{p}_{k}^{\prime }-\mathbf{p}_{j}-\mathbf{p}_{k})}{%
\sqrt{\omega _{jp_{j}^{\prime }}\omega _{kp_{k}^{\prime }}\omega
_{jp_{j}}\omega _{kp_{k}}}}\mathcal{M}_{s_{j}s_{k}s_{j}^{\prime
}s_{k}^{\prime }}^{\text{\textsl{Repulsive} }}(\mathbf{p}_{j},\mathbf{p}_{k},%
\mathbf{p}_{j}^{\prime },\mathbf{p}_{k}^{\prime })\Bigg\},  \label{21}
\end{eqnarray}
where $\displaystyle\underset{k=a}{\Sigma }^{\prime }u_{k}$ means $%
u_{a}+u_{a+2}+u_{a+4}+\cdots $. Our convention is that variables with odd
indices correspond to particles (e.g. $e^{-}$), and those with even indices
correspond to antiparticles (e.g. $e^{+}$).

The superscript notation $(j,k)$ in $\displaystyle\underset{i=1...n}{\prod }%
^{(j,k)}\,\delta ^{3}(\mathbf{p}_{i}^{\prime }-\mathbf{p}_{i})$ 
means that the terms with indices $j$ and $k$ are left out: 
\begin{equation}
\underset{i=1..n}{\prod }^{(j,k)}~\delta ^{3}(\mathbf{p}_{i}^{\prime }-%
\mathbf{p}_{i})=\prod_{i=1}^{j-1}\delta ^{3}(\mathbf{p}_{i}^{\prime }-%
\mathbf{p}_{i})\,\prod_{i=j+1}^{k-1}\delta ^{3}(\mathbf{p}_{i}^{\prime }-%
\mathbf{p}_{i})\,\prod_{i=k+1}^{n}\delta ^{3}(\mathbf{p}_{i}^{\prime }-%
\mathbf{p}_{i})=\frac{\prod_{i=1}^{n}\delta ^{3}(\mathbf{p}_{i}^{\prime }-%
\mathbf{p}_{i})}{\delta ^{3}(\mathbf{p}_{j}^{\prime }-\mathbf{p}%
_{j})\,\delta ^{3}(\mathbf{p}_{k}^{\prime }-\mathbf{p}_{k})},  \label{22}
\end{equation}
and similarly for $\underset{i=1...n}{\prod }^{(j,\text{ }k)}~\delta
_{s_{i}^{\prime }s_{i}}$ we have

\begin{equation}  \label{22a}
\underset{i=1...n}{\prod }^{(j,\text{ }k)}~\delta _{s_{i}^{\prime}s_{i}} =
\prod_{i=1}^{j-1}\delta_{s_{i}^{\prime}s_{i}}~
\prod_{i=j+1}^{k-1}\delta_{s_{i}^{\prime}s_{i}}~
\prod_{i=k+1}^{n}\delta_{s_{i}^{\prime}s_{i}} =\frac{\prod_{i=1}^{n}%
\delta_{s_{i}^{\prime}s_{i}}} {\delta_{s_{j}^{\prime}s_{j}}~\delta_{s_{k}^{%
\prime }s_{k}}}.
\end{equation}

For the case $n=2$, $\underset{i=1...n}{\prod }^{(j,\text{ }k)}~\delta ^{3}(%
\mathbf{p}_{i}^{\prime }-\mathbf{p}_{i})=1$ and $\underset{i=1...n}{\prod }%
^{(j,\text{ }k)}~\delta _{s_{i}^{\prime }s_{i}}=1$. But, for example, for $%
n=4$ (i.e. a four-body system), we have six terms. We write equations (\ref
{22}) and (\ref{22a}) explicitly for say, $j=2,$ $k=3$. 
\begin{eqnarray}
\underset{i=1...4}{\prod }^{(2,\text{ }3)}\delta ^{3}(\mathbf{p}_{i}^{\prime
}-\mathbf{p}_{i}) &=&\frac{\overset{4}{\underset{i=1}{\prod }}\delta ^{3}(%
\mathbf{p}_{i}^{\prime }-\mathbf{p}_{i})}{\delta ^{3}(\mathbf{p}_{2}^{\prime
}-\mathbf{p}_{2})\delta ^{3}(\mathbf{p}_{3}^{\prime }-\mathbf{p}_{3})}=\frac{%
\delta ^{3}(\mathbf{p}_{1}^{\prime }-\mathbf{p}_{1})\delta ^{3}(\mathbf{p}%
_{2}^{\prime }-\mathbf{p}_{2})\delta ^{3}(\mathbf{p}_{3}^{\prime }-\mathbf{p}%
_{3})\delta ^{3}(\mathbf{p}_{4}^{\prime }-\mathbf{p}_{4})}{\delta ^{3}(%
\mathbf{p}_{2}^{\prime }-\mathbf{p}_{2})\delta ^{3}(\mathbf{p}_{3}^{\prime }-%
\mathbf{p}_{3})},  \notag \\
&=&\delta ^{3}(\mathbf{p}_{1}^{\prime }-\mathbf{p}_{1})\delta ^{3}(\mathbf{p}%
_{4}^{\prime }-\mathbf{p}_{4}),  \label{ss1}
\end{eqnarray}
\begin{equation}
\underset{i=1...4}{\prod }^{(2,\text{ }3)}~\delta _{s_{i}^{\prime }s_{i}}=%
\frac{\overset{4}{\underset{i=1}{\prod }}\delta _{s_{i}^{\prime }s_{i}}}{%
\delta _{s_{2}^{\prime }s_{2}}~\delta _{s_{3}^{\prime }s_{3}}}=\frac{\delta
_{s_{1}^{\prime }s_{1}}\,\delta _{s_{2}^{\prime }s_{2}}\,\delta
_{s_{3}^{\prime }s_{3}}\,\delta _{s_{4}^{\prime }s_{4}}}{\delta
_{s_{2}^{\prime }s_{2}}~\delta _{s_{3}^{\prime }s_{3})}}=\delta
_{s_{1}^{\prime }s_{1}}\,\delta _{s_{4}^{\prime }s_{4}}.  \label{kk2}
\end{equation}

The expressions for $\mathcal{M}_{s_{j}s_{k}s_{j}^{\prime }s_{k}^{\prime }}^{%
\text{\textsl{Attractive} }}$, and $\mathcal{M}_{s_{j}s_{k}s_{j}^{\prime
}s_{k}^{\prime }}^{\text{\textsl{Repulsive} }}$ are as follows: 
\begin{eqnarray}
\mathcal{M}_{s_{j}s_{k}s_{j}^{\prime }s_{k}^{\prime }}^{\text{\textsl{%
Attractive} }}(\mathbf{p}_{j},\mathbf{p}_{k},\mathbf{p}_{j}^{\prime },%
\mathbf{p}_{k}^{\prime }) &=&\overline{u}_{j}(\mathbf{p}_{j}^{\prime
},s_{j}^{\prime })\gamma ^{\mu }u_{j}(\mathbf{p}_{j},s_{j})~[D_{\mu \nu
}(\omega _{jp_{j}^{\prime }}-\omega _{jp_{j}},\mathbf{p}_{j}^{\prime }-%
\mathbf{p}_{j})  \notag \\
&&+D_{\mu \nu }(\omega _{kp_{k}^{\prime }}-\omega _{kp_{k}},\mathbf{p}%
_{k}^{\prime }-\mathbf{p}_{k})]~\overline{v}_{k}\left( \mathbf{p}%
_{k},s_{k}\right) \gamma ^{\nu }v_{k}\left( \mathbf{p}_{k}^{\prime
},s_{k}^{\prime }\right) ,  \label{23}
\end{eqnarray}
if $j$ is odd (\textsl{e.g.} $e^{-}$) and $k$ is even (\textsl{e.g.} $\mu
^{+}$), and a similar expression, with $u$ replaced by $v,$ and $v$ replaced
by $u$ in equation (\ref{23}) if $j$ is even (\textsl{i.e.} $\mu ^{+}$) and $%
k$ is odd (\textsl{i.e.} $e^{-}$). The terms corresponding to one photon
exchange interactions among particles with same sign of charge are 
\begin{eqnarray}
\mathcal{M}_{s_{j}s_{k}s_{j}^{\prime }s_{k}^{\prime }}^{\text{\textsl{%
Repulsive} }}(\mathbf{p}_{j},\mathbf{p}_{k},\mathbf{p}_{j}^{\prime },\mathbf{%
p}_{k}^{\prime }) &=&\overline{u}_{j}(\mathbf{p}_{j}^{\prime },s_{j}^{\prime
})\gamma ^{\mu }u_{j}(\mathbf{p}_{j},s_{j})~[D_{\mu \nu }(\omega
_{jp_{j}^{\prime }}-\omega _{jp_{j}},\mathbf{p}_{j}^{\prime }-\mathbf{p}_{j})
\notag \\
&&+D_{\mu \nu }(\omega _{kp_{k}^{\prime }}-\omega _{kp_{k}},\mathbf{p}%
_{k}^{\prime }-\mathbf{p}_{k})]~\overline{u}_{k}\left( \mathbf{p}%
_{k}^{\prime },s_{k}^{\prime }\right) \gamma ^{\nu }u_{k}\left( \mathbf{p}%
_{k},s_{k}\right) ,  \label{24}
\end{eqnarray}
if $j$ and $k$ are both odd (\textsl{i.e.~} $e^{-}\mu ^{-}$) and a similar
expression, with $u$ replaced by $v$ in equation (\ref{24}) if $j$ and $k$
are both even (\textsl{i.e.~} $e^{+}\mu ^{+}$). %
Note that equations (\ref{23}) and (\ref{24}) correspond to one-photon
exchange Feynman diagrams between any two particles in the system, where (%
\ref{23}) applies to particles of opposite sign and (\ref{24}) to particles
of the same sign.

\vs1 The relativistic $n$-body wave equations for the coefficient functions $%
F_{s_{1}s_{2}...s_{n}}(\mathbf{p}_{1},...,\mathbf{p}_{n})$ of the trial
state (\ref{17}) or (\ref{18}) that follows from $\delta \langle \Psi_n |:{%
\hat{H}}-E:|\Psi_n \rangle _{t=0}=0,$ \thinspace\ is 
\begin{eqnarray}
&&F_{s_{1}s_{2}...s_{n}}(\mathbf{p}_{1},...,\mathbf{p}_{n}) \big[%
\omega_{1p_{1}}+\cdots +\omega _{np_{n}}-E\big] =\underset{s_{1}^{\prime
}...s_{n}^{\prime }}{\sum }\int d^{3}p_{1}^{\prime }...d^{3}p_{n}^{\prime
}~F_{s_{1}^{\prime }s_{2}^{\prime }...s_{n}^{\prime }}(\mathbf{p}%
_{1}^{\prime },...,\mathbf{p}_{n}^{\prime })  \notag \\
&&\Bigg\{{\overset{n-1}{\underset{j=1}{\sum }}}~{{\sum_{k=j+1}^{n}}}{%
\overset{\overset{}{\!\!^{^{\prime }}}}{}}~~\left[ \underset{i=1...n}{\prod }%
^{(j,k)}\delta _{s_{i}^{\prime }s_{i}}\right] ~\left[ \underset{i=1...n}{%
\prod }^{(j,k)}\delta ^{3}(\mathbf{p}_{i}^{\prime }-\mathbf{p}_{i})\right] 
\frac{m_{j}m_{k}q_{j}q_{k}}{2(2\pi )^{3}}\frac{\delta ^{3}(\mathbf{p}%
_{j}^{\prime }+\mathbf{p}_{k}^{\prime }-\mathbf{p}_{j}-\mathbf{p}_{k})}{%
\sqrt{\omega _{jp_{j}^{\prime }}\omega _{kp_{k}^{\prime }}\omega
_{jp_{j}}\omega _{kp_{k}}}}\times  \notag \\
&&\mathcal{M}_{s_{j}s_{k}s_{j}^{\prime }s_{k}^{\prime }}^{\text{\textsl{%
Attractive} }}(\mathbf{p}_{j},\mathbf{p}_{k},\mathbf{p}_{j}^{\prime },%
\mathbf{p}_{k}^{\prime })~~-{~~\overset{n-2}{\underset{j=1}{\sum }}~{%
\sum_{k=j+2}^{n}}}{\overset{\overset{}{\!\!^{^{\prime }}}}{}}~~\left[ 
\underset{i=1...n}{\prod }^{(j,k)}\delta _{s_{i}^{\prime }s_{i}}\right] %
\left[ \underset{i=1...n}{\prod }^{(j,k)}\delta ^{3}(\mathbf{p}_{i}^{\prime
}-\mathbf{p}_{i})\right] \frac{m_{j}m_{k}q_{j}q_{k}}{2(2\pi )^{3}}\times 
\notag \\
&&\frac{\delta ^{3}(\mathbf{p}_{j}^{\prime }+\mathbf{p}_{k}^{\prime }-%
\mathbf{p}_{j}-\mathbf{p}_{k})}{\sqrt{\omega _{jp_{j}^{\prime }}\omega
_{kp_{k}^{\prime }}\omega _{jp_{j}}\omega _{kp_{k}}}}\mathcal{M}%
_{s_{j}s_{k}s_{j}^{\prime }s_{k}^{\prime }}^{\text{\textsl{Repulsive} }}(%
\mathbf{p}_{j},\mathbf{p}_{k},\mathbf{p}_{j}^{\prime },\mathbf{p}%
_{k}^{\prime })\Bigg\},  \label{26}
\end{eqnarray}
where $\displaystyle\underset{k=a}{\Sigma }^{\prime }u_{k}$ means $%
u_{a}+u_{a+2}+u_{a+4}+\cdots $, as before.\smallskip

Equation (\ref{26}) is our main result. It is a relativistic momentum-space
equation for stationary (stable bound or scattering) states of a $n$-fermion
system, consisting of the same number of different-mass fermions and
antifermions if $n$ is even or with the number of particles one larger than
the number of antiparticles (or vice versa) if $n$ is odd, but with no
particle-antiparticle pairs. It is Salpeter-like (Schr\"{o}dinger-like) in
structure, with positive-energy solutions only, as can be seen by setting
the right-hand-side of eq. (\ref{26}) to zero, 
whereupon $E=\sum_{j}\omega _{jp_{j}}>0$ (as it must be given our use of the
``filled negative-energy'' vacuum $|0{\rangle }$, defined below eq. (\ref{14}%
)). 
In this respect eq. (\ref{26}) is different from many-fermion Dirac-like
equations or Bethe-Salpeter equations, which do have negative-energy
solutions (these, however, are generally disregarded in studies of two or
more body bound state systems).

One should note that the equations (\ref{26}) derived in the present article
are relativistic equations in which the kinematics of the $n$-fermion system
with arbitrary masses are treated exactly and the centre of mass motion is
taken into account without any approximation, that is the trial state (\ref
{17}) or (\ref{18}) is an eigenstate of the total momentum operator ${\hat{%
\mathbf{P}}}$ of the field theory with eigenvalue $\mathbf{Q}$ (see eq. (\ref
{15})). [$\mathbf{Q}$ can be taken to be zero in the rest frame of the
system.] Thus the mass polarization is built into the relativistic equations
since no assumptions about any of the particle masses being infinitely heavy
are made. There is no need for perturbative expansions in mass ratios.

The interaction kernels (relativistic momentum space potentials) in eq. (\ref
{26}) contain only tree-level Feynman diagrams (cf. (\ref{23}), (\ref{24})),
that is, one-quantum exchange, including retardation effects. This means
that physical effects to $O(\alpha ^{4})$ only are contained in eq. (\ref{26}%
). (This has been shown explicitly for the two and three fermion systems in
earlier studies, \cite{TD0} for positronium, \cite{TD1} for muonium and \cite
{Mu08} for $e^+ e^- e^-$ and $\mu^+ e^- e^-$, as discussed in more detail in
section 3 below.) 
To include effects beyond $O(\alpha ^{4})$ requires the use of more
elaborate approximations than just the simplest trial states (\ref{17}) and (%
\ref{18}), as has been illustrated on the relatively simple case of
positronium (cf. ref. \cite{TDH07}). Alternatively, one could ``cheat'' by
simply adding matrix elements corresponding to higher-order (loop) diagrams
to the kernels (\ref{23}) and (\ref{24}).

%
To help understand the content of the approximations inherent in equation (%
\ref{26}) it is useful to consider its non-relativistic limit, \textit{i.e.\;%
} when %
$\mathbf{p}^{2}/m^{2}$ $<<1$. In this limit $D_{\mu \nu }$ in\ the
expressions for $\mathcal{M}_{s_{j}s_{k}s_{j}^{\prime }s_{k}^{\prime }}^{%
\text{\textsl{Attractive} }}$and $\mathcal{M}_{s_{j}s_{k}s_{j}^{\prime
}s_{k}^{\prime }}^{\text{\textsl{Repulsive} }}$ reduce to $D_{00}=1/\mathbf{%
|q-p|}^{2}$ (and zero otherwise) for the attractive and repulsive terms.
Thus, in the non relativistic limit, 
\begin{equation}
\mathcal{\tilde{M}}_{s_{j}s_{k}s_{j}^{\prime }s_{k}^{\prime }}^{\text{%
\textsl{Attractive} }}=2\,\overline{u}(0,s_{j}^{\prime })\gamma
^{0}u(0,s_{j})\frac{1}{\mathbf{|p}_{j}^{\prime }\mathbf{-p}_{j}\mathbf{|}^{2}%
}\overline{v}\left( 0,s_{k}\right) \gamma ^{0}v\left( 0,s_{k}^{\prime
}\right) =\frac{2\delta _{s_{j}s_{j}^{\prime }}\delta _{s_{k}s_{k}^{\prime }}%
}{\mathbf{|p}_{j}^{\prime }\mathbf{-p}_{j}\mathbf{|}^{2}},  \label{34}
\end{equation}
and 
\begin{equation}
\mathcal{\tilde{M}}_{s_{j}s_{k}s_{j}^{\prime }s_{k}^{\prime }}^{\text{%
\textsl{Repulsive} }}=2\,\overline{u}(0,s_{j}^{\prime })\gamma ^{0}u(0,s_{j})%
\frac{1}{\mathbf{|p}_{j}^{\prime }\mathbf{-p}_{j}\mathbf{|}^{2}}\overline{u}%
\left( 0,s_{k}^{\prime }\right) \gamma ^{0}u\left( 0,s_{k}\right) =\frac{%
2\delta _{s_{j}s_{j}^{\prime }}\delta _{s_{k}s_{k}^{\prime }}}{\mathbf{|p}%
_{j}^{\prime }\mathbf{-p}_{j}\mathbf{|}^{2}}.  \label{35}
\end{equation}

For arbitrary $n$, the coordinate-space equation, obtained by Fourier
transformation %
\begin{equation}
F_{s_{1}s_{2}...s_{n}}(\mathbf{p}_{1},...,\mathbf{p}_{n})=\frac{1}{(2\pi
)^{3n/2}}\int d^{3}x_{1}...d^{3}x_{n}~\Psi _{s_{1}s_{2}...s_{n}}(\mathbf{x}%
_{1},...,\mathbf{x}_{n})\,e^{-i(\mathbf{p}_{1}\cdot \mathbf{x}_{1}+\cdots +%
\mathbf{p}_{n}\cdot \mathbf{x}_{n})},  \label{38}
\end{equation}
of the non-relativistic limit of eq. (\ref{26}), is as expected, the $n$%
-body Schr\"{o}dinger equation, 
\begin{equation}
\left[ -{\sum_{i=1}^{n}}\frac{1}{2m_{i}}{\nabla _{i}^{2}}-\epsilon -{%
\overset{n-1}{\underset{j=1}{\sum }}}~{{\sum_{k=j+1}^{n}}}{\overset{\overset{%
}{\!\!^{^{\prime }}}}{}}\frac{\alpha _{jk}}{|\mathbf{x}_{j}-\mathbf{x}_{k}|}%
~+~{\overset{n-2}{\underset{j=1}{\sum }}~{\sum_{k=j+2}^{n}}}{\overset{%
\overset{}{\!\!^{^{\prime }}}}{}}\frac{\alpha _{jk}}{|\mathbf{x}_{j}-\mathbf{%
x}_{k}|}\,\right] \,\Psi _{s_{1}...s_{n}}(\mathbf{x}_{1},...,\mathbf{x}%
_{n})~=0,  \label{45}
\end{equation}
where $\epsilon =E-[m_{1}+m_{2}+\cdots +m_{n}]$, $\displaystyle\alpha _{jk}=%
\frac{q_{j}q_{k}}{4\pi }$ are the coupling constants, and $\displaystyle%
\underset{k=a}{\Sigma }^{\prime }u_{k}$ means $u_{a}+u_{a+2}+u_{a+4}+\cdots $%
.

In the non-relativistic limit, we see that the same equations are obtained
for all $\Psi _{s_{1}...s_{n}}$ (or, equivalently, for all $%
F_{s_{1}...s_{n}} $), hence the spin and space parts of the non-relativistic
wave functions separate, and we can write $
\Psi _{s_{1}s_{2}...s_{n}}(\mathbf{x}_{1},...,\mathbf{x}_{n})= \Lambda
_{s_{1}s_{2}...s_{n}} \,\Psi (\mathbf{x}_{1},...,\mathbf{x}_{n}). ~ $%
The $\Lambda _{s_{1}s_{2}...s_{n}}$ are spin coefficients that must be
chosen so that $\displaystyle \sum_{s_1...s_n} F_{s_{1}...s_{n}}$ is the
appropriate $n$-body angular-momentum eigenstate. 
%

The variationally obtained $n$-body relativistic equations (\ref{26}) (or
non-relativistic equations (\ref{45}) for $n>2$) are not analytically
solvable, hence approximate, usually variational, solutions must be
obtained. Thus, for any $n$ and any state, one can obtain an approximate
variational solution for the energy $E_{n}$ and the wave-function of the $n$%
-body system by replacing $F_{s_{1}s_{2}...s_{n}}(\mathbf{p}_{1},...,\mathbf{%
p}_{n})$ with analytic functions containing adjustable features (parameters)
to compute the energy expectation value (see Eqs. (\ref{19})-(\ref{21})): 
\begin{equation}
E_{n}=\frac{\langle \Psi _{n}|:{\hat{H}}:|\Psi _{n}\rangle }{\langle \Psi
_{n}|\Psi _{n}\rangle }.  \label{5.1a}
\end{equation}
Optimal values of the adjustable features (parameters) of the trial wave
functions correspond to the minimum values of (\ref{5.1a}). This minimum
principle allows for a systematic improvement of the approximate
(variational) solutions. %
\vs1

\section{Two, three and four-body examples:}

The relativistic equations (\ref{26}) have been previously derived and
solved approximately for the $n=2$, two-body cases (like muonium, $\mu
^{+}e^{-}$, ref. \cite{TD1}) and the $n=3$, three-body cases (like $\mu
^{+}e^{-}e^{-}$, ref. \cite{Mu08}). It was shown that results, correct to $%
O(\alpha ^{4})$, are obtained for the energies of all bound states of these
systems. We shall recount some details.

Thus, for the two-body problem $(e^- \mu^+)$, for which the trial state (\ref
{17}) is 
\begin{equation}
|\Psi _{2}\rangle =\underset{s_{1}s_{2}}{\sum }\int
d^{3}p_{1}d^{3}p_{2}~F_{s_{1}s_{2}}(\mathbf{p}_{1},\mathbf{p}%
_{2})~b_{1}^{\dagger }(\mathbf{p}_{1},s_{1})d_{2}^{\dagger }(\mathbf{p}%
_{2},s_{2})|0\rangle ,  \label{27Bis}
\end{equation}
the wave equation (\ref{26}) becomes: 
\begin{eqnarray}
F_{s_{1}s_{2}}(\mathbf{p}_{1},\mathbf{p}_{2})~[\omega _{1p_{1}}+\omega
_{2p_{2}} &-&E]=\frac{q_{1}q_{2}m_{1}m_{2}}{2(2\pi )^{3}}\underset{%
s_{1}^{\prime }s_{2}^{\prime }}{\sum }\int d^{3}p_{1}^{\prime
}d^{3}p_{2}^{\prime }~F_{s_{1}^{\prime }s_{2}^{\prime }}(\mathbf{p}%
_{1}^{\prime },\mathbf{p}_{2}^{\prime })\frac{\delta ^{3}(\mathbf{p}%
_{1}^{\prime }+\mathbf{p}_{2}^{\prime }-\mathbf{p}_{1}-\mathbf{p}_{2})}{%
\sqrt{\omega _{1p_{1}^{\prime }}\omega _{2p_{2}^{\prime }}\omega
_{1p_{1}}\omega _{2p2}}}  \notag \\
&&\times \left[ \mathcal{M}_{s_{1}s_{2}s_{1}^{\prime }s_{2}^{\prime }}^{%
\text{\textsl{Attractive} }}(\mathbf{p}_{1},\mathbf{p}_{2},\mathbf{p}%
_{1}^{\prime },\mathbf{p}_{2}^{\prime })\right],  \label{27}
\end{eqnarray}
\begin{eqnarray}
\mathcal{M}_{s_{1}s_{2}s_{1}^{\prime }s_{2}^{\prime }}^{\text{\textsl{%
Attractive} }}(\mathbf{p}_{1},\mathbf{p}_{2},\mathbf{p}_{1}^{\prime },%
\mathbf{p}_{2}^{\prime }) &=&\overline{u}_{1}(\mathbf{p}_{1}^{\prime
},s_{1}^{\prime })\gamma ^{\mu }u_{1}(\mathbf{p}_{1},s_{1})~[D_{\mu \nu
}(\omega _{1p_{1}^{\prime }}-\omega _{1p_{1}},\mathbf{p}_{1}^{\prime }-%
\mathbf{p}_{1})  \notag \\
&&+D_{\mu \nu }(\omega _{2p_{2}^{\prime }}-\omega _{2p_{2}},\mathbf{p}%
_{2}^{\prime }-\mathbf{p}_{2})]~\overline{v}_{2}\left( \mathbf{p}%
_{2},s_{2}\right) \gamma ^{\nu }v_{2}\left( \mathbf{p}_{2}^{\prime
},s_{2}^{\prime }\right) ,  \label{28}
\end{eqnarray}
(Of course, there is no repulsive term for $(e^- \mu^+)$ case.) Equations (%
\ref{27}) and (\ref{28}) were derived previously \cite{TD1}. We mention, in
passing, that for a system consisting a particle and antiparticle of equal
mass (like positronium) an additional virtual annihilation interaction term, 
$\mathcal{M}_{s_{j}s_{k}s_{j}^{\prime }s_{k}^{\prime }}^{\text{\textsl{%
Annihilation} }}$ arises in Eq. (\ref{27}) as shown in \cite{TD0}: 
\begin{eqnarray}
\mathcal{M}_{s_{1}s_{2}s_{1}^{\prime }s_{2}^{\prime }}^{\text{\textsl{%
Annihilation} }}(\mathbf{p}_{1},\mathbf{p}_{2},\mathbf{p}_{1}^{\prime },%
\mathbf{p}_{2}^{\prime }) &=&\overline{u}(\mathbf{p}_{1}^{\prime
},s_{1}^{\prime })\gamma ^{\mu }v\left( \mathbf{p}_{2}^{\prime
},s_{2}^{\prime }\right) ~[D_{\mu \nu }(\omega _{p_{1}^{\prime }}+\omega
_{p_{2}^{\prime }},\mathbf{p}_{1}^{\prime }+\mathbf{p}_{2}^{\prime })  \notag
\\
&&+D_{\mu \nu }(-\omega _{p_{1}}-\omega _{p_{2}},-\mathbf{p}_{1}-\mathbf{p}%
_{2})]~\overline{v}\left( \mathbf{p}_{2},s_{2}\right) \gamma ^{\nu }u(%
\mathbf{p}_{1},s_{1}).  \label{29}
\end{eqnarray}
The variational two-fermion wave equations for muonium like systems were
solved approximately by Terekidi and Darewych \cite{TD1}; their results were
shown to be in agreement with other calculations and in good agreement with
the observed muonium spectrum to O($\alpha ^{4}$) (as well as that for
hydrogen and muonic hydrogen).

The non-relativistic limit of the wave equation (\ref{27}) for the two-body
system $(e^{-}\mu ^{+})$ is 
\begin{eqnarray}
&&F_{s_{1}s_{2}}(\mathbf{p}_{1},\mathbf{p}_{2})~\left[ \frac{\mathbf{p}%
_{1}^{2}}{2m_{1}}+\frac{\mathbf{p}_{2}^{2}}{2m_{2}}-\epsilon _{2}\right] 
\notag \\
&=&\frac{q_{1}q_{2}}{(2\pi )^{3}}\,\underset{s_{1}^{\prime }s_{2}^{\prime }}{%
\sum }\int d^{3}p_{1}^{\prime }d^{3}p_{2}^{\prime }~F_{s_{1}^{\prime
}s_{2}^{\prime }}(\mathbf{p}_{1}^{\prime },\mathbf{p}_{2}^{\prime })~\delta
^{3}(\mathbf{p}_{1}^{\prime }+\mathbf{p}_{2}^{\prime }-\mathbf{p}_{1}-%
\mathbf{p}_{2})\left[ \frac{\delta _{s_{1}s_{1}^{\prime }}\delta
_{s_{2}s_{2}^{\prime }}}{\mathbf{|p}_{1}^{\prime }\mathbf{-p}_{1}\mathbf{|}%
^{2}}\right] ,  \label{37}
\end{eqnarray}
where $\epsilon _{2}=E-[m_{1}+m_{2}]$. The coordinate-space form of eq. (\ref
{37}), is, of course, equation (\ref{45}) with $n=2$, namely the expected
Schr\"{o}dinger equation, 
\begin{equation}
\left[ -{\sum_{i=1}^{2}\nabla _{i}^{2}}\frac{1}{2m_{i}}-\frac{\alpha }{|%
\mathbf{x}_{1}-\mathbf{x}_{2}|}-\epsilon _{2}\right] \Psi _{s_{1}s_{2}}(%
\mathbf{x}_{1},\mathbf{x}_{2})=0,  \label{38a}
\end{equation}
where $\displaystyle\alpha =\frac{q_{1}q_{2}}{4\pi }$ is the usual
fine-structure constant and $|q_{1}|=|q_{2}|=|e|,$ where $e$ is the
elementary charge.

For the general case of three constituents with different masses, systems
like $(m_{1}^{-}m_{2}^{+}m_{3}^{-})$, with the trial state (cf. eq. (\ref{18}%
)), 
\begin{equation}
|\Psi _{3}\rangle =\underset{s_{1}s_{2}s_{3}}{\sum }\int
d^{3}p_{1}d^{3}p_{2}d^{3}p_{3}~F_{s_{1}s_{2}s_{3}}(\mathbf{p}_{1},\mathbf{p}%
_{2},\mathbf{p}_{3})b_{1}^{\dagger }(\mathbf{p}_{1},s_{1})d_{2}^{\dagger }(%
\mathbf{p}_{2},s_{2})b_{3}^{\dagger }(\mathbf{p}_{3},s_{3})|0\rangle ,
\label{40a}
\end{equation}
the three-body wave equation is (\ref{26}) with $n=3$, namely 
\begin{eqnarray}
&&F_{s_{1}s_{2}s_{3}}(\mathbf{p}_{1},\mathbf{p}_{2},\mathbf{p}_{3})[\omega
_{1p_{1}}+\omega _{2p_{2}}+\omega _{3p_{3}}-E]=\underset{s_{1}^{\prime
}s_{2}^{\prime }s_{3}^{\prime }}{\sum }\int d^{3}p_{1}^{\prime
}d^{3}p_{2}^{\prime }d^{3}p_{3}^{\prime }~F_{s_{1}^{\prime }s_{2}^{\prime
}s_{3}^{\prime }}(\mathbf{p}_{1}^{\prime },\mathbf{p}_{2}^{\prime },\mathbf{p%
}_{3}^{\prime })\times  \notag \\
&&\Bigg\{\frac{m_{1}m_{2}q_{1}q_{2}}{2(2\pi )^{3}}\mathcal{M}%
_{s_{1}s_{2}s_{1}^{\prime }s_{2}^{\prime }}^{\text{\textsl{Attractive} }}(%
\mathbf{p}_{1},\mathbf{p}_{2},\mathbf{p}_{1}^{\prime },\mathbf{p}%
_{2}^{\prime })\delta _{s_{3}^{\prime }s_{3}}\delta ^{3}(\mathbf{p}%
_{3}^{\prime }-\mathbf{p}_{3})\frac{\delta ^{3}(\mathbf{p}_{1}^{\prime }+%
\mathbf{p}_{2}^{\prime }-\mathbf{p}_{1}-\mathbf{p}_{2})}{\sqrt{\omega
_{1p_{1}^{\prime }}\omega _{2p_{2}^{\prime }}\omega _{1p_{1}}\omega _{2p_{2}}%
}}  \notag \\
&&+\frac{m_{2}m_{3}q_{2}q_{3}}{2(2\pi )^{3}}\mathcal{M}_{s_{2}s_{3}s_{2}^{%
\prime }s_{3}^{\prime }}^{\text{\textsl{Attractive} }}(\mathbf{p}_{2},%
\mathbf{p}_{3},\mathbf{p}_{2}^{\prime },\mathbf{p}_{3}^{\prime })\delta
_{s_{1}^{\prime }s_{1}}\delta ^{3}(\mathbf{p}_{1}^{\prime }-\mathbf{p}_{1})%
\frac{\delta ^{3}(\mathbf{p}_{2}^{\prime }+\mathbf{p}_{3}^{\prime }-\mathbf{p%
}_{2}-\mathbf{p}_{3})}{\sqrt{\omega _{2p_{2}^{\prime }}\omega
_{3p_{3}^{\prime }}\omega _{2p_{2}}\omega _{3p_{3}}}}  \notag \\
&&-\frac{m_{1}m_{3}q_{1}q_{3}}{2(2\pi )^{3}}\mathcal{M}_{s_{1}s_{3}s_{1}^{%
\prime }s_{3}^{\prime }}^{\text{\textsl{Repulsive} }}(\mathbf{p}_{1},\mathbf{%
p}_{3},\mathbf{p}_{1}^{\prime },\mathbf{p}_{3}^{\prime })~\delta
_{s_{2}^{\prime }s_{2}}~\delta ^{3}(\mathbf{p}_{2}^{\prime }-\mathbf{p}_{2})%
\frac{\delta ^{3}(\mathbf{p}_{1}^{\prime }+\mathbf{p}_{3}^{\prime }-\mathbf{p%
}_{1}-\mathbf{p}_{3})}{\sqrt{\omega _{1p_{1}^{\prime }}\omega
_{3p_{3}^{\prime }}\omega _{1p_{1}}\omega _{3p_{3}}}}\Bigg\}.  \label{40b}
\end{eqnarray}
The expressions for $\mathcal{M}_{s_{j}s_{k}s_{j}^{\prime }s_{k}^{\prime }}^{%
\text{\textsl{Attractive} }}$and $\mathcal{M}_{s_{j}s_{k}s_{j}^{\prime
}s_{k}^{\prime }}^{\text{\textsl{Repulsive} }}$ are given in equations (\ref
{23}) and (\ref{24}), respectively. Note that for H$^{-}$, and the muonium
negative ion, Mu$^{-}$ ($e^{-}\mu ^{+}e^{-})$, we have the same wave
equation (\ref{40b}) but with two of the constituents with equal masses.
Note again that $|q_{1}|=|q_{2}|=|q_{3}|=|e|,$ where $e$ is the elementary
charge. As already mentioned, approximate solutions of the relativistic
three fermion equation (\ref{40b}) for some systems are presented in ref. 
\cite{Mu08}. They are in agreement with other calculations and with
experimental results to $O(\alpha^4)$.

In the non-relativistic limit, eq. (\ref{40b}) reduces to the following: 
\begin{eqnarray}
&&F_{s_{1}s_{2}s_{3}}(\mathbf{p}_{1},\mathbf{p}_{2},\mathbf{p}_{3})\left[ 
\frac{\mathbf{p}_{1}^{2}}{2m_{1}}+\frac{\mathbf{p}_{2}^{2}}{2m_{2}}+\frac{%
\mathbf{p}_{3}^{2}}{2m_{3}}-\epsilon _{3}\right] =\underset{s_{1}^{\prime
}s_{2}^{\prime }s_{3}^{\prime }}{\sum }\int d^{3}p_{1}^{\prime
}d^{3}p_{2}^{\prime }d^{3}p_{3}^{\prime }~F_{s_{1}^{\prime }s_{2}^{\prime
}s_{3}^{\prime }}(\mathbf{p}_{1}^{\prime },\mathbf{p}_{2}^{\prime },\mathbf{p%
}_{3}^{\prime })  \notag \\
&&\Bigg\{\frac{q_{1}q_{2}}{(2\pi )^{3}}\frac{\delta _{s_{1}s_{1}^{\prime
}}\delta _{s_{2}s_{2}^{\prime }}\delta _{s_{3}^{\prime }s_{3}}}{\mathbf{|p}%
_{1}^{\prime }\mathbf{-p}_{1}\mathbf{|}^{2}}~\delta ^{3}(\mathbf{p}%
_{3}^{\prime }-\mathbf{p}_{3})\delta ^{3}(\mathbf{p}_{1}^{\prime }+\mathbf{p}%
_{2}^{\prime }-\mathbf{p}_{1}-\mathbf{p}_{2})+\frac{q_{2}q_{3}}{(2\pi )^{3}}%
\frac{\delta _{s_{1}^{\prime }s_{1}}\delta _{s_{2}s_{2}^{\prime }}\delta
_{s_{3}^{\prime }s_{3}}}{\mathbf{|p}_{2}^{\prime }\mathbf{-p}_{2}\mathbf{|}%
^{2}}\delta ^{3}(\mathbf{p}_{1}^{\prime }-\mathbf{p}_{1})\times  \notag \\
&&\delta ^{3}(\mathbf{p}_{2}^{\prime }+\mathbf{p}_{3}^{\prime }-\mathbf{p}%
_{2}-\mathbf{p}_{3})-\frac{q_{1}q_{3}}{(2\pi )^{3}}\frac{\delta
_{s_{1}^{\prime }s_{1}}\delta _{s_{2}s_{2}^{\prime }}\delta _{s_{3}^{\prime
}s_{3}}}{\mathbf{|p}_{1}^{\prime }\mathbf{-p}_{1}\mathbf{|}^{2}}~\delta ^{3}(%
\mathbf{p}_{2}^{\prime }-\mathbf{p}_{2})\delta ^{3}(\mathbf{p}_{1}^{\prime }+%
\mathbf{p}_{3}^{\prime }-\mathbf{p}_{1}-\mathbf{p}_{3})\Bigg\},  \label{41}
\end{eqnarray}
where $\epsilon _{3}=E-[m_{1}+m_{2}+m_{3}]$. In the coordinate space, this
becomes the three-body Schr\"{o}dinger equation (\ref{45}), with $n=3$,
namely 
\begin{equation}
\Bigg\{-\left( {\sum_{i=1}^{3}}\frac{1}{2m_{i}}{\nabla _{i}^{2}}\right)
-\epsilon _{3}-\frac{\alpha _{12}}{|\mathbf{x}_{1}-\mathbf{x}_{2}|}-\frac{%
\alpha _{23}}{|\mathbf{x}_{2}-\mathbf{x}_{3}|}+\frac{\alpha _{13}}{|\mathbf{x%
}_{1}-\mathbf{x}_{3}|}\Bigg\}\Psi _{s_{1}s_{2}s_{3}}(\mathbf{x}_{1},\mathbf{x%
}_{2},\mathbf{x}_{3})=0,  \label{A3}
\end{equation}
where $\displaystyle\alpha _{jk}=\frac{q_{j}q_{k}}{4\pi }$, as expected.

For the four-body case of various ``flavors'' $%
(m_{1}^{-}m_{2}^{+}m_{3}^{-}m_{4}^{+})$ the relativistic equation is (\ref
{26}) with $n=4$, that is 
\begin{eqnarray}
&&F_{s_{1}s_{2}s_{3}s_{4}}(\mathbf{p}_{1},\mathbf{p}_{2},\mathbf{p}_{3},%
\mathbf{p}_{4})\,\big[\omega _{1p_{1}}+\omega _{2p_{2}}+\omega
_{3p_{3}}+\omega _{4p_{4}}-E\big]  \label{55} \\
&=&\underset{s_{1}^{\prime }s_{2}^{\prime }s_{3}^{\prime }s_{4}^{\prime }}{%
\sum }\int d^{3}p_{1}^{\prime }d^{3}p_{2}^{\prime }d^{3}p_{3}^{\prime
}d^{3}p_{4}^{\prime }~F_{s_{1}^{\prime }s_{2}^{\prime }s_{3}^{\prime
}s_{4}^{\prime }}(\mathbf{p}_{1}^{\prime },\mathbf{p}_{2}^{\prime },\mathbf{p%
}_{3}^{\prime },\mathbf{p}_{4}^{\prime })\times  \notag \\
&&\Bigg\{\frac{m_{1}m_{2}q_{1}q_{2}}{2(2\pi )^{3}}\mathcal{M}%
_{s_{1}s_{2}s_{1}^{\prime }s_{2}^{\prime }}^{\text{\textsl{Attractive} }}(%
\mathbf{p}_{1},\mathbf{p}_{2},\mathbf{p}_{1}^{\prime },\mathbf{p}%
_{2}^{\prime })~\delta _{s_{3}^{\prime }s_{3}}\delta _{s_{4}^{\prime
}s_{4}}~\delta ^{3}(\mathbf{p}_{3}^{\prime }-\mathbf{p}_{3})\delta ^{3}(%
\mathbf{p}_{4}^{\prime }-\mathbf{p}_{4})\frac{\delta ^{3}(\mathbf{p}%
_{1}^{\prime }+\mathbf{p}_{2}^{\prime }-\mathbf{p}_{1}-\mathbf{p}_{2})}{%
\sqrt{\omega _{1p_{1}^{\prime }}\omega _{2p_{2}^{\prime }}\omega
_{1p_{1}}\omega _{2p_{2}}}}  \notag \\
&&+\frac{m_{2}m_{3}q_{2}q_{3}}{2(2\pi )^{3}}\mathcal{M}_{s_{2}s_{3}s_{2}^{%
\prime }s_{3}^{\prime }}^{\text{\textsl{Attractive} }}(\mathbf{p}_{2},%
\mathbf{p}_{3},\mathbf{p}_{2}^{\prime },\mathbf{p}_{3}^{\prime })~\delta
_{s_{1}^{\prime }s_{1}}\delta _{s_{4}^{\prime }s_{4}}~\delta ^{3}(\mathbf{p}%
_{1}^{\prime }-\mathbf{p}_{1})\delta ^{3}(\mathbf{p}_{4}^{\prime }-\mathbf{p}%
_{4})\frac{\delta ^{3}(\mathbf{p}_{2}^{\prime }+\mathbf{p}_{3}^{\prime }-%
\mathbf{p}_{2}-\mathbf{p}_{3})}{\sqrt{\omega _{2p_{2}^{\prime }}\omega
_{3p_{3}^{\prime }}\omega _{2p_{2}}\omega _{3p_{3}}}}  \notag \\
&&+\frac{m_{3}m_{4}q_{3}q_{4}}{2(2\pi )^{3}}\mathcal{M}_{s_{3}s_{4}s_{3}^{%
\prime }s_{4}^{\prime }}^{\text{\textsl{Attractive} }}(\mathbf{p}_{3},%
\mathbf{p}_{4},\mathbf{p}_{3}^{\prime },\mathbf{p}_{4}^{\prime })~\delta
_{s_{1}^{\prime }s_{1}}\delta _{s_{2}^{\prime }s_{2}}~\delta ^{3}(\mathbf{p}%
_{1}^{\prime }-\mathbf{p}_{1})\delta ^{3}(\mathbf{p}_{2}^{\prime }-\mathbf{p}%
_{2})\frac{\delta ^{3}(\mathbf{p}_{3}^{\prime }+\mathbf{p}_{4}^{\prime }-%
\mathbf{p}_{3}-\mathbf{p}_{4})}{\sqrt{\omega _{3p_{3}^{\prime }}\omega
_{4p_{4}^{\prime }}\omega _{3p_{3}}\omega _{4p_{4}}}}  \notag \\
&&+\frac{m_{1}m_{4}q_{1}q_{4}}{2(2\pi )^{3}}\mathcal{M}_{s_{1}s_{4}s_{1}^{%
\prime }s_{4}^{\prime }}^{\text{\textsl{Atrractive}}}(\mathbf{p}_{1},\mathbf{%
p}_{4},\mathbf{p}_{1}^{\prime },\mathbf{p}_{4}^{\prime })~\delta
_{s_{2}^{\prime }s_{2}}\delta _{s_{3}^{\prime }s_{3}}~\delta ^{3}(\mathbf{p}%
_{2}^{\prime }-\mathbf{p}_{2})\delta ^{3}(\mathbf{p}_{3}^{\prime }-\mathbf{p}%
_{3})\frac{\delta ^{3}(\mathbf{p}_{1}^{\prime }+\mathbf{p}_{4}^{\prime }-%
\mathbf{p}_{1}-\mathbf{p}_{4})}{\sqrt{\omega _{1p_{1}^{\prime }}\omega
_{4p_{4}^{\prime }}\omega _{1p_{1}}\omega _{4p_{4}}}}  \notag \\
&&-\frac{m_{1}m_{3}q_{1}q_{3}}{2(2\pi )^{3}}\mathcal{M}_{s_{1}s_{3}s_{1}^{%
\prime }s_{3}^{\prime }}^{\text{\textsl{Repulsive} }}(\mathbf{p}_{1},\mathbf{%
p}_{3},\mathbf{p}_{1}^{\prime },\mathbf{p}_{3}^{\prime })~\delta
_{s_{2}^{\prime }s_{2}}\delta _{s_{4}^{\prime }s_{4}}~\delta ^{3}(\mathbf{p}%
_{2}^{\prime }-\mathbf{p}_{2})\delta ^{3}(\mathbf{p}_{4}^{\prime }-\mathbf{p}%
_{4})\frac{\delta ^{3}(\mathbf{p}_{1}^{\prime }+\mathbf{p}_{3}^{\prime }-%
\mathbf{p}_{1}-\mathbf{p}_{3})}{\sqrt{\omega _{1p_{1}^{\prime }}\omega
_{3p_{3}^{\prime }}\omega _{1p_{1}}\omega _{3p_{3}}}}  \notag \\
&&-\frac{m_{2}m_{4}q_{2}q_{4}}{2(2\pi )^{3}}\mathcal{M}_{s_{2}s_{4}s_{2}^{%
\prime }s_{4}^{\prime }}^{\text{\textsl{Repulsive} }}(\mathbf{p}_{2},\mathbf{%
p}_{4},\mathbf{p}_{2}^{\prime },\mathbf{p}_{4}^{\prime })~\delta
_{s_{1}^{\prime }s_{1}}\delta _{s_{3}^{\prime }s_{3}}~\delta ^{3}(\mathbf{p}%
_{1}^{\prime }-\mathbf{p}_{1})\delta ^{3}(\mathbf{p}_{3}^{\prime }-\mathbf{p}%
_{3})\frac{\delta ^{3}(\mathbf{p}_{2}^{\prime }+\mathbf{p}_{4}^{\prime }-%
\mathbf{p}_{2}-\mathbf{p}_{4})}{\sqrt{\omega _{2p_{2}^{\prime }}\omega
_{4p_{4}^{\prime }}\omega _{2p_{2}}\omega _{4p_{4}}}}\Bigg\},  \notag
\end{eqnarray}
The expressions for $\mathcal{M}_{s_{j}s_{k}s_{j}^{\prime }s_{k}^{\prime }}^{%
\text{\textsl{Attractive} }}$ and $\mathcal{M}_{s_{j}s_{k}s_{j}^{\prime
}s_{k}^{\prime }}^{\text{\textsl{Repulsive} }}$ are given in equations (\ref
{23}) and (\ref{24}). No (approximate) solutions of the relativistic
four-fermion equation (\ref{55}) have been obtained to date. The
non-relativistic limit of (\ref{55}) is, of course, equation (\ref{45}) with 
$n=4$.

One can, analogously, write out the explicit expression for the relativistic 
$n$-fermion equation (\ref{26}) for $n=5$ or larger.

\section{Concluding remarks.}

The solution of the non-relativistic $n$-body system is a difficult problem
for $n>3$; all the more so for the relativistic counterpart. Theoretical
investigation of non-relativistic four-body systems interacting through
Coulombic potentials have been discussed in several works and studies have
been done regarding the existence of bound-states of such systems, including
the domain of stability in the space of inverse masses \cite{Varga2005}.
Some exotic, non-relativistic five-body systems have been investigated by
Mezei et al. \cite{Mitroy2001}, who studied the stability of a number of
five-body systems using stochastic variational method (SVM). However, much
remains to be done in the field of relativistic equations (including QED
effects) for atomic $n$-body systems with $n\geq 3$, particularly, systems
of different masses. Equation (\ref{26}) can be used to calculate
relativistic effects, though the effort required is considerable even in the
case of weak binding, when perturbation theory with respect to (approximate)
non-relativistic solutions is applicable.

One should note that the systems considered in this paper are basically
composed of fundamental fermions, such as electron ($e$), muon ($\mu $), or
tauon ($\tau $), and their corresponding antifermions. During the early days
of quantum mechanics a suitable set of coordinates and basis states for the
three-body problem was proposed by Hylleraas \cite{He} , and it was used to
calculate the ground state energy of the helium atom. At that time (1929)
the proton was considered to be a fundamental (``point'') particle (protons
are now known to have three-quark plus gluon substructure). Around half a
century after the work of Hylleraas \cite{He}-\cite{He1}, some theoretical
and experimental advances have been made and extensive high-precision
calculations became feasible (see Drake \cite{Drake1}-\cite{Drake33}).
Moreover, we can also mention the following example, which illustrates that
some unresolved QED problems are under investigation currently. One of the
intriguing questions which remain to be answered in bound-state quantum
electrodynamics is related to the discrepancy between the theoretical and
experimental values for the Lamb shift in ionized helium, or He$^{+}$, which
is a hydrogen like atomic system with a nuclear charge number Z = 2; the
current status of the subject {is discussed} in ref. \cite{Drake4}.

The situation is different for exotic atoms or ions that are ``pure'' QED
systems, such as the positronium negative ion (Ps$^{-}$) or the muonium
negative ion (Mu$^{-}$), bound only by electromagnetic interactions. To our
knowledge, the binding energy of Ps$^{-}$ or Mu$^{-}$ has not been measured
to date. The only four-body exotic system that is a ``pure'' QED system and
which has\textbf{\ }been observed is the positronium molecule \cite{Cassidy}%
. However, the experimental value of the binding energy has not as yet been
obtained for Ps$_{2}$ . The situation is even worse for systems of $n>3$
fermions of various mass; for example, $\mu ^{+}e^{-}\mu ^{+}e^{-}$ has not
even been observed as yet, and to our knowledge there is no relativistic or
QED study of this system (Mu$_{2}$). Therefore, we can say that we are still
in the early stage of study of these exotic pure QED systems. 
Clearly, the study of pure QED systems, bound only by electromagnetic
interactions, is of fundamental interest 
as was shown for the positronium ($e^{-}e^{+}$) system in some recent
experiments \cite{Cassidy1}-\cite{Cassidy2}.

To sum up, we have derived relativistic equations (\ref{26}) for systems of $%
n\geq 2$ fermions of various mass. The relativistic kinematics are included
exactly in equations (\ref{26}), but the interactions contain tree-level
interactions only (one photon exchange, and virtual annihilation in the case
of pairs), \textit{i.e.} they are incomplete beyond O($\alpha ^{4}$). To
calculate effects beyond this order, the matrix elements $\mathcal{M}$ in (%
\ref{26}) can be augmented by higher order (loop) diagram contributions, as
is done in the Bethe-Salpeter formalism 
(beyond the ladder approximation). %

Because our method of deriving the equations (including the interactions) is
variational, the description of 
$n$ fermion system can be improved systematically by using more elaborate $n$%
-body trial states than (\ref{17}) and (\ref{18}), as was done for the
two-body Ps system \cite{TDH07}. 
Variational trial states, such as (\ref{17}) and (\ref{18}), can be
generalized in various ways. Thus, the single Fock trial state $|\Psi
_{2}\rangle $ (\ref{27Bis}) can be replaced by a superposition of two (or
more) Fock states (as was done for scalar models \cite{E-RD06}-\cite{M1}),
including states that accommodate virtual pairs. We illustrate this on the
two-fermion $(\mu ^{+}\mu ^{-})$ system. The single Fock trial state $|\Psi
_{t}\rangle =|\Psi _{2}\rangle $, where\textbf{\ 
\begin{equation}
|\Psi _{2}\rangle =\underset{s_{1}s_{2}}{\sum }\int
d^{3}p_{1}d^{3}p_{2}~F_{s_{1}s_{2}}(p_{1},p_{2})~b_{\mu }^{\dagger
}(p_{1},s_{1})d_{\mu }^{\dagger }(p_{2},s_{2})|0\rangle ,  \label{JD1}
\end{equation}
}can be generalized to $|\Psi _{t}\rangle =|\Psi _{2}\rangle +|\Psi
_{4}\rangle $, where\textbf{\ 
\begin{equation}
|\Psi _{4}\rangle =\underset{s_{1}...s_{4}}{\sum }\int
d^{3}p_{1}...d^{3}p_{4}~G_{s_{1}...s_{4}}(p_{1}...p_{4})~b_{\mu }^{\dagger
}(p_{1},s_{1})d_{\mu }^{\dagger }(p_{2},s_{2})b_{e}^{\dagger
}(p_{3},s_{3})d_{e}^{\dagger }(p_{4},s_{4})|0\rangle ,  \label{JD2}
\end{equation}
}where $b_{e}^{\dagger }$ and $d_{e}^{\dagger }$ are electron and positron
creation operators. The variational principle (\ref{16}) would then lead to
coupled, multi-dimensional integral equations for the channel wave-functions 
$F$ and $G$. The channel function $G$, evidently, accommodates an
electron-positron pair, which is virtual when the system energy domain is $%
E<2\,m_{\mu }$. Clearly the coupled multidimensional equations for $F$ and $%
G $ can only be solved approximately, say variationally, but this is a
tedious calculation that will be left for the future.


%
%

It should be mentioned that numerical calculations have been done in order
to solve approximately relativistic wave equations of $n$-body systems of
scalar particles and antiparticle of various masses (for example ref. \cite
{M2011}, for $n=3$). However, determining approximate solutions of
relativistic wave equations for $n$-body systems ($n\geq 3$) of fermions and
antifermions of various masses is considerably more difficult and remains a
challenging task. As mentioned previously, it is not possible to solve the
relativistic $n$-body equations derived in this paper analytically.
Therefore, approximate (\textsl{i.e.}, numerical, variational or
perturbative) solutions must be sought for various cases of interest. The
stochastic variational and quantum Monte Carlo methods are popular among
methods for the computation of complicated exotic systems \cite{Varga}-\cite
{Sh}. In Ref. \cite{BL}, the possible production of systems such as true
muonium ($\mu ^{+}\mu ^{-}$), true tauonium ($\tau ^{+}\tau ^{-}$), and
``mu-tauonium'' ($\mu ^{\pm }\tau ^{\mp }$) has been discussed. The
discovery of ($\mu ^{+}\mu ^{-}$) in future will herald a new chapter for
the observation of much more difficult systems such as ($\mu ^{\pm }\tau
^{\mp }$). In any case, experiments on other exotic atoms or molecules,
though difficult, will be undertaken in future even though it may not be the
near future.

Lastly, we wish to point out that it would be of interest to \textbf{a}pply
the variational method and the Hamiltonian formalism used in the present
work to QCD systems in order to derive relativistic, momentum space integral
equations for systems consisting of $n$ quarks and anti quarks, interacting
via gluon exchange, as has been attempted in a previous paper for a
quark-antiquark system \cite{DiLeoDar}.


\end{document}